\documentclass[RNAAS]{aastex631}
\usepackage[utf8]{inputenc}

\shorttitle{esdT candidate W0523}
\shortauthors{Samuel J. Goodman}

\begin{document}

\title{WISEA J052305.94-015356.1: A New EsdT Candidate}

\correspondingauthor{Samuel J. Goodman}
\email{samuel.goodman@zohomail.eu}

\author[0000-0003-2236-2320]{Samuel J. Goodman}
\affiliation{Amateur Astronomer, Kent, UK; samuel.goodman@zohomail.eu}

\begin{abstract}
I present WISEA J052305.94-015356.1 as a new candidate extremely metal-poor T subdwarf (esdT), based on its distinctive infrared colours and high proper motion ($\sim500\ $mas/yr). Spectroscopic follow-up is now needed to confirm it is a member of this newly discovered class of substellar objects.
\end{abstract}

\section{Introduction} \label{sec:intro}
Recently, \citet{Schneider_2020} reported the discovery of what are likely the first two extreme subdwarfs of the T spectral class (esdT; [Fe/H] $\leqslant -1$ dex, Teff $\lesssim 1400$ K); WISEA J041451.67-585456.7 (W0414) and WISEA J181006.18-101000.5 (W1810). A key feature of the two suspected esdTs are their unusual near to mid-infrared colours; some colours are reminiscent of late L dwarfs, while others appear like those of mid T dwarfs \citep{Schneider_2020}. A useful colour-colour plot to identify these objects is the W1-W2 vs. J-W2 diagram, commonly used in brown dwarf studies. The currently known esdTs appear clustered in a locus at W1$-$W2 colours which would typically indicate an early-to-mid T dwarf, however they lie much redder than the T dwarf sequence in J-W2. For example, \citet{Meisner_2021} in their study chose colour cuts of $1.1$ mag $<$ W1-W2 $< 1.75$ mag and J-W2 $> 3$ mag; A region that appears relatively unpopulated by known objects in the ultracool dwarf regime. \citet{Meisner_2021} demonstrate that the existence of significantly low-metallicity brown dwarfs in this locus is predicted by synthetic photometric colours derived from new "LOWZ" atmospheric models which they present. 

\begin{figure}
  \includegraphics[width=0.8\paperwidth]{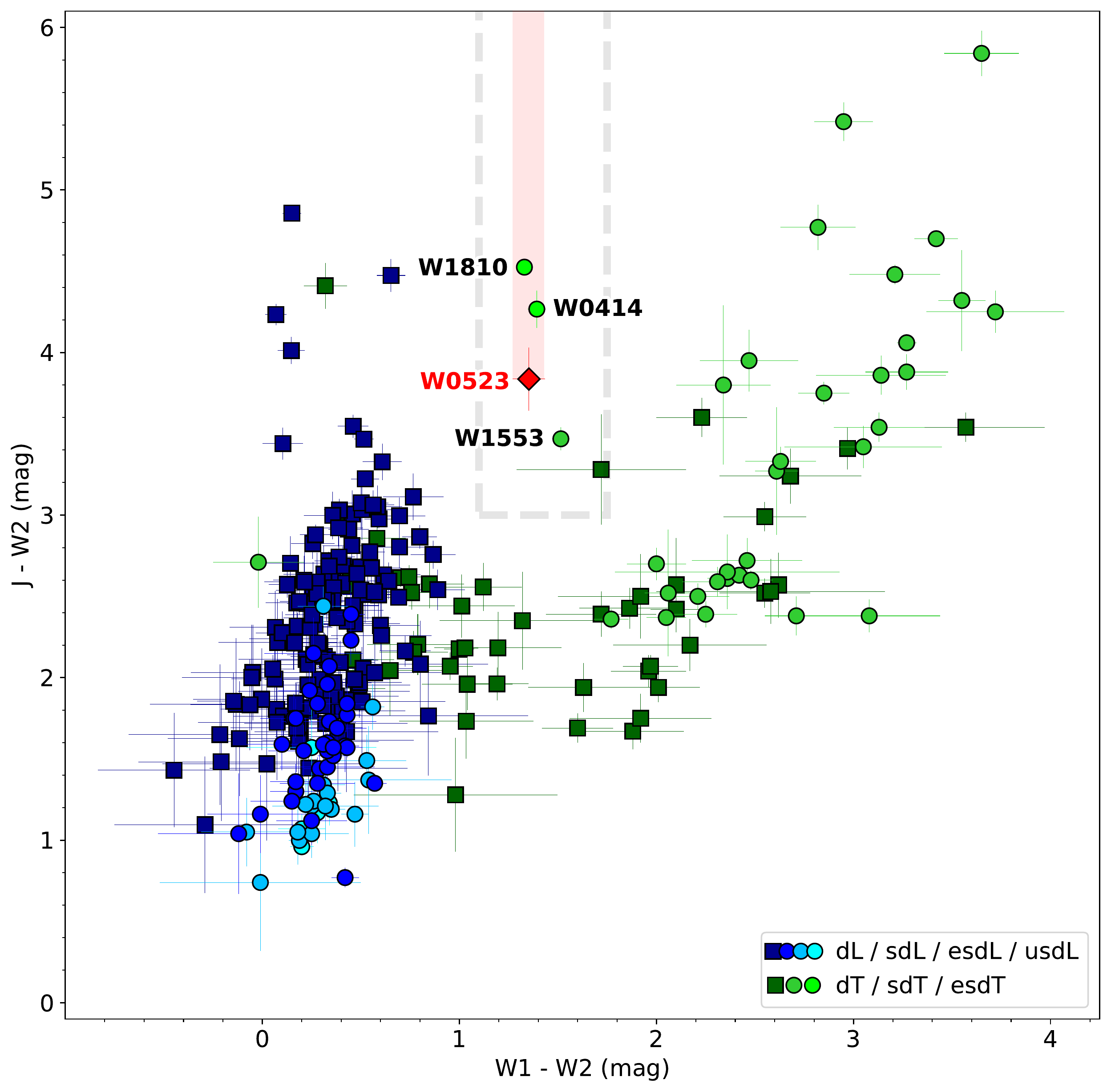}
  \caption{W1-W2 vs. J-W2 colour-colour diagram. W0523, the two known esdTs, as well as the sdT interloper W1553 \citep{Meisner_2021} are labelled. The gray dashed box represents the esdT colour selection from \citet{Meisner_2021}. The pale red solid box represents the region the candidate would occupy should the J-band counterpart be unrelated. L and T dwarfs are shown with dark blue and dark green squares respectively, while metal-poor subclasses are displayed in circles with progressively lighter shades for their respective spectral type. L and T subdwarfs come from \citet{Zhang_2018}, \citet{Zhang_2019a}, and \citet{Zhang_2019b}, while L and T dwarfs are sourced from \citet{Burningham_2013} and \citet{Marocco_2015}.}
  \label{fig:colourcolour}
\end{figure}

\section{Candidate} \label{sec:candidate}
I present a new candidate esdT, WISEA J052305.94-015356.1 (W0523), discovered through visual inspection of sources in the CatWISE 2020 catalogue \citep{Marocco_2021} displaying significant proper motion. The source (J052306.42-015355.4 in the CatWISE 2020 catalogue) is visibly seen to move in animations of unWISE coadded images \citep{Lang_2014,Meisner_2018} taken as part of the WISE \citep{Wright_2010} and NEOWISE \citep{Mainzer_2011,Mainzer_2014} surveys in the WiseView web-tool \citep{Caselden_2018}. Its motion-compensated W1-W2 colour ($1.35 \pm 0.08$ mag) is in the range commonly associated with mid T dwarfs, but has a significantly redder J-W2 colour (at least $3.84 \pm 0.19$ mag; see next paragraph for further details) by almost two magnitudes and hence falling within the esdT colour cuts of \citet{Meisner_2021}. The source possesses high proper motion ($\mu = 513 \pm 53$ mas/yr) which reduces the likelihood that it is a misidentified extra-galactic contaminant, as well as being a potential indicator of thick disk or halo membership. The W2 reduced proper motion (H$_{W2}$ = m$_{W2}$ + 5 + 5 * log10($\mu$), where $\mu$ is in arcseconds; \citealt{Luyten_1922}) of W0523 is $19.46 \pm 0.23$; high but not as excessive as the sdT interloper W1553. This is however similar to esdT W0414's H$_{W2}$ of 19.47 \citep{Meisner_2021}. The W1-W2 colour and polynomial relations from \citep{Kirkpatrick_2021} would suggest a T5 dwarf spectral type, though both the spectral fitting of known esdTs as well as the LOWZ model evolutionary tracks imply that if it is extremely metal-poor then 'normal' T dwarf relations may not hold.

The VISTA Hemisphere Survey (VHS) \citep{McMahon_2013} provides comparatively deep imaging in J and Ks-bands at this sky position, obtained in 2009 and so before the first WISE epoch. Two VHS DR6 catalogue sources accessed through the Vista Science Archive are located within 10 $\arcsec$ of the AllWISE position; one (ID 473519603563) located 1.515 $\arcsec$ away and the other (ID 473519603589) 2.626 $\arcsec$ away. The more distant source is detected in both the Sloan Digital Sky Survey (SDSS) \citep{Alam_2015} and Pan-STARRS \citep{Chambers_2016} with colours suggesting a background object rather than an ultracool dwarf. Unlike its neighbour, the closest near-infrared source has no optical counterpart in either SDSS or Pan-STARRS and is only detected in the J-band image. The closest source has an automated classification as a galaxy in the VHS survey, however for very faint sources that are near the limit of detection such as this one (J $= 19.748 \pm 0.186$ mag) the automated determination may be unreliable. The large uncertainty in the WISE position of candidate W0523 due to its faint magnitude (W2 $= 15.912 \pm 0.055$ mag) and WISE's low angular resolution make it difficult to be certain the WISE candidate and this source are truly related. The 2.626 $\arcsec$ background located beneath the candidate may be contaminating its WISE flux, moving the centroid to lower declination, which may in part account for the separation between the closest source and the candidate's AllWISE position. I therefore tentatively assign the closest source as the J-band counterpart to our esdT candidate. Even if the closest source is not in fact a counterpart, this implies non-detection of the candidate in J-band and an even more red J-W2 colour than I determine here. As a result, the interpretation that the object is an esdT candidate would not change if the J-band source is found to be an unassociated background source. I show the location of W0523 on a W1-W2 vs. J-W2 diagram in Figure~\ref{fig:colourcolour}.

\begin{acknowledgments}
This publication makes use of data products from the Wide-field Infrared Survey Explorer, which is a joint project of the University of California, Los Angeles, and the Jet Propulsion Laboratory/California Institute of Technology, and NEOWISE which is a project of the Jet Propulsion Laboratory/California Institute of Technology. WISE and NEOWISE are funded by the National Aeronautics and Space Administration. Based on observations obtained as part of the VISTA Hemisphere Survey, ESO Program, 179.A-2010 (PI: McMahon).

\end{acknowledgments}

\bibliography{esdt_Bib.bib}{}

\begin{thebibliography}{}
\expandafter\ifx\csname natexlab\endcsname\relax\def\natexlab#1{#1}\fi
\providecommand{\url}[1]{\href{#1}{#1}}
\providecommand{\dodoi}[1]{doi:~\href{http://doi.org/#1}{\nolinkurl{#1}}}
\providecommand{\doeprint}[1]{\href{http://ascl.net/#1}{\nolinkurl{http://ascl.net/#1}}}
\providecommand{\doarXiv}[1]{\href{https://arxiv.org/abs/#1}{\nolinkurl{https://arxiv.org/abs/#1}}}

\bibitem[{{Alam} {et~al.}(2015){Alam}, {Albareti}, {Allende Prieto}, {Anders},
  {Anderson}, {Anderton}, {Andrews}, {Armengaud}, {Aubourg}, {Bailey}, {Basu},
  {Bautista}, {Beaton}, {Beers}, {Bender}, {Berlind}, {Beutler}, {Bhardwaj},
  {Bird}, {Bizyaev}, {Blake}, {Blanton}, {Blomqvist}, {Bochanski}, {Bolton},
  {Bovy}, {Shelden Bradley}, {Brandt}, {Brauer}, {Brinkmann}, {Brown},
  {Brownstein}, {Burden}, {Burtin}, {Busca}, {Cai}, {Capozzi}, {Carnero
  Rosell}, {Carr}, {Carrera}, {Chambers}, {Chaplin}, {Chen}, {Chiappini},
  {Chojnowski}, {Chuang}, {Clerc}, {Comparat}, {Covey}, {Croft}, {Cuesta},
  {Cunha}, {da Costa}, {Da Rio}, {Davenport}, {Dawson}, {De Lee}, {Delubac},
  {Deshpande}, {Dhital}, {Dutra-Ferreira}, {Dwelly}, {Ealet}, {Ebelke},
  {Edmondson}, {Eisenstein}, {Ellsworth}, {Elsworth}, {Epstein}, {Eracleous},
  {Escoffier}, {Esposito}, {Evans}, {Fan}, {Fern{\'a}ndez-Alvar}, {Feuillet},
  {Filiz Ak}, {Finley}, {Finoguenov}, {Flaherty}, {Fleming}, {Font-Ribera},
  {Foster}, {Frinchaboy}, {Galbraith-Frew}, {Garc{\'\i}a},
  {Garc{\'\i}a-Hern{\'a}ndez}, {Garc{\'\i}a P{\'e}rez}, {Gaulme}, {Ge},
  {G{\'e}nova-Santos}, {Georgakakis}, {Ghezzi}, {Gillespie}, {Girardi},
  {Goddard}, {Gontcho}, {Gonz{\'a}lez Hern{\'a}ndez}, {Grebel}, {Green},
  {Grieb}, {Grieves}, {Gunn}, {Guo}, {Harding}, {Hasselquist}, {Hawley},
  {Hayden}, {Hearty}, {Hekker}, {Ho}, {Hogg}, {Holley-Bockelmann}, {Holtzman},
  {Honscheid}, {Huber}, {Huehnerhoff}, {Ivans}, {Jiang}, {Johnson},
  {Kinemuchi}, {Kirkby}, {Kitaura}, {Klaene}, {Knapp}, {Kneib}, {Koenig},
  {Lam}, {Lan}, {Lang}, {Laurent}, {Le Goff}, {Leauthaud}, {Lee}, {Lee},
  {Licquia}, {Liu}, {Long}, {L{\'o}pez-Corredoira}, {Lorenzo-Oliveira},
  {Lucatello}, {Lundgren}, {Lupton}, {Mack}, {Mahadevan}, {Maia}, {Majewski},
  {Malanushenko}, {Malanushenko}, {Manchado}, {Manera}, {Mao}, {Maraston},
  {Marchwinski}, {Margala}, {Martell}, {Martig}, {Masters}, {Mathur},
  {McBride}, {McGehee}, {McGreer}, {McMahon}, {M{\'e}nard}, {Menzel},
  {Merloni}, {M{\'e}sz{\'a}ros}, {Miller}, {Miralda-Escud{\'e}}, {Miyatake},
  {Montero-Dorta}, {More}, {Morganson}, {Morice-Atkinson}, {Morrison},
  {Mosser}, {Muna}, {Myers}, {Nandra}, {Newman}, {Neyrinck}, {Nguyen},
  {Nichol}, {Nidever}, {Noterdaeme}, {Nuza}, {O'Connell}, {O'Connell},
  {O'Connell}, {Ogando}, {Olmstead}, {Oravetz}, {Oravetz}, {Osumi}, {Owen},
  {Padgett}, {Padmanabhan}, {Paegert}, {Palanque-Delabrouille}, {Pan},
  {Parejko}, {P{\^a}ris}, {Park}, {Pattarakijwanich}, {Pellejero-Ibanez},
  {Pepper}, {Percival}, {P{\'e}rez-Fournon}, {P{\'e}rez-R{\`a}fols},
  {Petitjean}, {Pieri}, {Pinsonneault}, {Porto de Mello}, {Prada}, {Prakash},
  {Price-Whelan}, {Protopapas}, {Raddick}, {Rahman}, {Reid}, {Rich}, {Rix},
  {Robin}, {Rockosi}, {Rodrigues}, {Rodr{\'\i}guez-Torres}, {Roe}, {Ross},
  {Ross}, {Rossi}, {Ruan}, {Rubi{\~n}o-Mart{\'\i}n}, {Rykoff},
  {Salazar-Albornoz}, {Salvato}, {Samushia}, {S{\'a}nchez}, {Santiago},
  {Sayres}, {Schiavon}, {Schlegel}, {Schmidt}, {Schneider}, {Schultheis},
  {Schwope}, {Sc{\'o}ccola}, {Scott}, {Sellgren}, {Seo}, {Serenelli}, {Shane},
  {Shen}, {Shetrone}, {Shu}, {Silva Aguirre}, {Sivarani}, {Skrutskie},
  {Slosar}, {Smith}, {Sobreira}, {Souto}, {Stassun}, {Steinmetz}, {Stello},
  {Strauss}, {Streblyanska}, {Suzuki}, {Swanson}, {Tan}, {Tayar}, {Terrien},
  {Thakar}, {Thomas}, {Thomas}, {Thompson}, {Tinker}, {Tojeiro}, {Troup},
  {Vargas-Maga{\~n}a}, {Vazquez}, {Verde}, {Viel}, {Vogt}, {Wake}, {Wang},
  {Weaver}, {Weinberg}, {Weiner}, {White}, {Wilson}, {Wisniewski},
  {Wood-Vasey}, {Ye`che}, {York}, {Zakamska}, {Zamora}, {Zasowski}, {Zehavi},
  {Zhao}, {Zheng}, {Zhou}, {Zhou}, {Zou}, \& {Zhu}}]{Alam_2015}
{Alam}, S., {Albareti}, F.~D., {Allende Prieto}, C., {et~al.} 2015, \apjs, 219,
  12, \dodoi{10.1088/0067-0049/219/1/12}

\bibitem[{{Burningham} {et~al.}(2013){Burningham}, {Cardoso}, {Smith},
  {Leggett}, {Smart}, {Mann}, {Dhital}, {Lucas}, {Tinney}, {Pinfield}, {Zhang},
  {Morley}, {Saumon}, {Aller}, {Littlefair}, {Homeier}, {Lodieu}, {Deacon},
  {Marley}, {van Spaandonk}, {Baker}, {Allard}, {Andrei}, {Canty}, {Clarke},
  {Day-Jones}, {Dupuy}, {Fortney}, {Gomes}, {Ishii}, {Jones}, {Liu},
  {Magazz{\'u}}, {Marocco}, {Murray}, {Rojas-Ayala}, \&
  {Tamura}}]{Burningham_2013}
{Burningham}, B., {Cardoso}, C.~V., {Smith}, L., {et~al.} 2013, \mnras, 433,
  457, \dodoi{10.1093/mnras/stt740}

\bibitem[{{Caselden} {et~al.}(2018){Caselden}, {Westin}, {Meisner}, {Kuchner},
  \& {Colin}}]{Caselden_2018}
{Caselden}, D., {Westin}, Paul, I., {Meisner}, A., {Kuchner}, M., \& {Colin},
  G. 2018, {WiseView: Visualizing motion and variability of faint WISE
  sources}.
\newblock \doeprint{1806.004}

\bibitem[{{Chambers} {et~al.}(2016){Chambers}, {Magnier}, {Metcalfe},
  {Flewelling}, {Huber}, {Waters}, {Denneau}, {Draper}, {Farrow}, {Finkbeiner},
  {Holmberg}, {Koppenhoefer}, {Price}, {Rest}, {Saglia}, {Schlafly}, {Smartt},
  {Sweeney}, {Wainscoat}, {Burgett}, {Chastel}, {Grav}, {Heasley}, {Hodapp},
  {Jedicke}, {Kaiser}, {Kudritzki}, {Luppino}, {Lupton}, {Monet}, {Morgan},
  {Onaka}, {Shiao}, {Stubbs}, {Tonry}, {White}, {Ba{\~n}ados}, {Bell},
  {Bender}, {Bernard}, {Boegner}, {Boffi}, {Botticella}, {Calamida},
  {Casertano}, {Chen}, {Chen}, {Cole}, {Deacon}, {Frenk}, {Fitzsimmons},
  {Gezari}, {Gibbs}, {Goessl}, {Goggia}, {Gourgue}, {Goldman}, {Grant},
  {Grebel}, {Hambly}, {Hasinger}, {Heavens}, {Heckman}, {Henderson}, {Henning},
  {Holman}, {Hopp}, {Ip}, {Isani}, {Jackson}, {Keyes}, {Koekemoer}, {Kotak},
  {Le}, {Liska}, {Long}, {Lucey}, {Liu}, {Martin}, {Masci}, {McLean}, {Mindel},
  {Misra}, {Morganson}, {Murphy}, {Obaika}, {Narayan}, {Nieto-Santisteban},
  {Norberg}, {Peacock}, {Pier}, {Postman}, {Primak}, {Rae}, {Rai}, {Riess},
  {Riffeser}, {Rix}, {R{\"o}ser}, {Russel}, {Rutz}, {Schilbach}, {Schultz},
  {Scolnic}, {Strolger}, {Szalay}, {Seitz}, {Small}, {Smith}, {Soderblom},
  {Taylor}, {Thomson}, {Taylor}, {Thakar}, {Thiel}, {Thilker}, {Unger},
  {Urata}, {Valenti}, {Wagner}, {Walder}, {Walter}, {Watters}, {Werner},
  {Wood-Vasey}, \& {Wyse}}]{Chambers_2016}
{Chambers}, K.~C., {Magnier}, E.~A., {Metcalfe}, N., {et~al.} 2016, arXiv
  e-prints, arXiv:1612.05560.
\newblock \doarXiv{1612.05560}

\bibitem[{{Kirkpatrick} {et~al.}(2021){Kirkpatrick}, {Gelino}, {Faherty},
  {Meisner}, {Caselden}, {Schneider}, {Marocco}, {Cayago}, {Smart},
  {Eisenhardt}, {Kuchner}, {Wright}, {Cushing}, {Allers}, {Bardalez Gagliuffi},
  {Burgasser}, {Gagn{\'e}}, {Logsdon}, {Martin}, {Ingalls}, {Lowrance},
  {Abrahams}, {Aganze}, {Gerasimov}, {Gonzales}, {Hsu}, {Kamraj}, {Kiman},
  {Rees}, {Theissen}, {Ammar}, {Andersen}, {Beaulieu}, {Colin}, {Elachi},
  {Goodman}, {Gramaize}, {Hamlet}, {Hong}, {Jonkeren}, {Khalil}, {Martin},
  {Pendrill}, {Pumphrey}, {Rothermich}, {Sainio}, {Stenner}, {Tanner},
  {Th{\'e}venot}, {Voloshin}, {Walla}, {W{\k{e}}dracki}, \& {Backyard Worlds:
  Planet 9 Collaboration}}]{Kirkpatrick_2021}
{Kirkpatrick}, J.~D., {Gelino}, C.~R., {Faherty}, J.~K., {et~al.} 2021, \apjs,
  253, 7, \dodoi{10.3847/1538-4365/abd107}

\bibitem[{{Lang}(2014)}]{Lang_2014}
{Lang}, D. 2014, \aj, 147, 108, \dodoi{10.1088/0004-6256/147/5/108}

\bibitem[{{Luyten}(1922)}]{Luyten_1922}
{Luyten}, W.~J. 1922, Lick Observatory Bulletin, 336, 135

\bibitem[{{Mainzer} {et~al.}(2011){Mainzer}, {Bauer}, {Grav}, {Masiero},
  {Cutri}, {Dailey}, {Eisenhardt}, {McMillan}, {Wright}, {Walker}, {Jedicke},
  {Spahr}, {Tholen}, {Alles}, {Beck}, {Brandenburg}, {Conrow}, {Evans},
  {Fowler}, {Jarrett}, {Marsh}, {Masci}, {McCallon}, {Wheelock}, {Wittman},
  {Wyatt}, {DeBaun}, {Elliott}, {Elsbury}, {Gautier}, {Gomillion}, {Leisawitz},
  {Maleszewski}, {Micheli}, \& {Wilkins}}]{Mainzer_2011}
{Mainzer}, A., {Bauer}, J., {Grav}, T., {et~al.} 2011, \apj, 731, 53,
  \dodoi{10.1088/0004-637X/731/1/53}

\bibitem[{{Mainzer} {et~al.}(2014){Mainzer}, {Bauer}, {Cutri}, {Grav},
  {Masiero}, {Beck}, {Clarkson}, {Conrow}, {Dailey}, {Eisenhardt}, {Fabinsky},
  {Fajardo-Acosta}, {Fowler}, {Gelino}, {Grillmair}, {Heinrichsen}, {Kendall},
  {Kirkpatrick}, {Liu}, {Masci}, {McCallon}, {Nugent}, {Papin}, {Rice},
  {Royer}, {Ryan}, {Sevilla}, {Sonnett}, {Stevenson}, {Thompson}, {Wheelock},
  {Wiemer}, {Wittman}, {Wright}, \& {Yan}}]{Mainzer_2014}
{Mainzer}, A., {Bauer}, J., {Cutri}, R.~M., {et~al.} 2014, \apj, 792, 30,
  \dodoi{10.1088/0004-637X/792/1/30}

\bibitem[{{Marocco} {et~al.}(2015){Marocco}, {Jones}, {Day-Jones}, {Pinfield},
  {Lucas}, {Burningham}, {Zhang}, {Smart}, {Gomes}, \& {Smith}}]{Marocco_2015}
{Marocco}, F., {Jones}, H.~R.~A., {Day-Jones}, A.~C., {et~al.} 2015, \mnras,
  449, 3651, \dodoi{10.1093/mnras/stv530}

\bibitem[{{Marocco} {et~al.}(2021){Marocco}, {Eisenhardt}, {Fowler},
  {Kirkpatrick}, {Meisner}, {Schlafly}, {Stanford}, {Garcia}, {Caselden},
  {Cushing}, {Cutri}, {Faherty}, {Gelino}, {Gonzalez}, {Jarrett}, {Koontz},
  {Mainzer}, {Marchese}, {Mobasher}, {Schlegel}, {Stern}, {Teplitz}, \&
  {Wright}}]{Marocco_2021}
{Marocco}, F., {Eisenhardt}, P. R.~M., {Fowler}, J.~W., {et~al.} 2021, \apjs,
  253, 8, \dodoi{10.3847/1538-4365/abd805}

\bibitem[{{McMahon} {et~al.}(2013){McMahon}, {Banerji}, {Gonzalez}, {Koposov},
  {Bejar}, {Lodieu}, {Rebolo}, \& {VHS Collaboration}}]{McMahon_2013}
{McMahon}, R.~G., {Banerji}, M., {Gonzalez}, E., {et~al.} 2013, The Messenger,
  154, 35

\bibitem[{{Meisner} {et~al.}(2018){Meisner}, {Lang}, \&
  {Schlegel}}]{Meisner_2018}
{Meisner}, A.~M., {Lang}, D., \& {Schlegel}, D.~J. 2018, \aj, 156, 69,
  \dodoi{10.3847/1538-3881/aacbcd}

\bibitem[{{Meisner} {et~al.}(2021){Meisner}, {Schneider}, {Burgasser},
  {Marocco}, {Line}, {Faherty}, {Kirkpatrick}, {Caselden}, {Kuchner}, {Gelino},
  {Gagne}, {Theissen}, {Gerasimov}, {Aganze}, {Hsu}, {Wisniewski}, {Casewell},
  {Bardalez Gagliuffi}, {Logsdon}, {Eisenhardt}, {Allers}, {Debes}, {Allen},
  {Stevnbak Andersen}, {Goodman}, {Gramaize}, {Martin}, {Sainio}, {Cushing},
  {Backyard Worlds}, {:}, \& {Planet 9 Collaboration}}]{Meisner_2021}
{Meisner}, A.~M., {Schneider}, A.~C., {Burgasser}, A.~J., {et~al.} 2021, arXiv
  e-prints, arXiv:2106.01387.
\newblock \doarXiv{2106.01387}

\bibitem[{{Schneider} {et~al.}(2020){Schneider}, {Burgasser}, {Gerasimov},
  {Marocco}, {Gagn{\'e}}, {Goodman}, {Beaulieu}, {Pendrill}, {Rothermich},
  {Sainio}, {Kuchner}, {Caselden}, {Meisner}, {Faherty}, {Mamajek}, {Hsu},
  {Greco}, {Cushing}, {Kirkpatrick}, {Bardalez-Gagliuffi}, {Logsdon}, {Allers},
  {Debes}, \& {Backyard Worlds: Planet 9 Collaboration}}]{Schneider_2020}
{Schneider}, A.~C., {Burgasser}, A.~J., {Gerasimov}, R., {et~al.} 2020, \apj,
  898, 77, \dodoi{10.3847/1538-4357/ab9a40}

\bibitem[{Wright {et~al.}(2010)Wright, Eisenhardt, Mainzer, Ressler, Cutri,
  Jarrett, Kirkpatrick, Padgett, McMillan, Skrutskie, Stanford, Cohen, Walker,
  Mather, Leisawitz, Gautier, McLean, Benford, Lonsdale, Blain, Mendez, Irace,
  Duval, Liu, Royer, Heinrichsen, Howard, Shannon, Kendall, Walsh, Larsen,
  Cardon, Schick, Schwalm, Abid, Fabinsky, Naes, \& Tsai}]{Wright_2010}
Wright, E.~L., Eisenhardt, P. R.~M., Mainzer, A.~K., {et~al.} 2010, The
  Astronomical Journal, 140, 1868, \dodoi{10.1088/0004-6256/140/6/1868}

\bibitem[{{Zhang}(2019)}]{Zhang_2019b}
{Zhang}, Z. 2019, \mnras, 489, 1423, \dodoi{10.1093/mnras/stz2196}

\bibitem[{{Zhang} {et~al.}(2019){Zhang}, {Burgasser}, {G{\'a}lvez-Ortiz},
  {Lodieu}, {Zapatero Osorio}, {Pinfield}, \& {Allard}}]{Zhang_2019a}
{Zhang}, Z.~H., {Burgasser}, A.~J., {G{\'a}lvez-Ortiz}, M.~C., {et~al.} 2019,
  \mnras, 486, 1260, \dodoi{10.1093/mnras/stz777}

\bibitem[{{Zhang} {et~al.}(2018){Zhang}, {Galvez-Ortiz}, {Pinfield},
  {Burgasser}, {Lodieu}, {Jones}, {Mart{\'\i}n}, {Burningham}, {Homeier},
  {Allard}, {Zapatero Osorio}, {Smith}, {Smart}, {L{\'o}pez Mart{\'\i}},
  {Marocco}, \& {Rebolo}}]{Zhang_2018}
{Zhang}, Z.~H., {Galvez-Ortiz}, M.~C., {Pinfield}, D.~J., {et~al.} 2018,
  \mnras, 480, 5447, \dodoi{10.1093/mnras/sty2054}

\end{thebibliography}
\bibliographystyle{aasjournal}

\end{document}